\documentclass[aps,prl,twocolumn,superscriptaddress,groupedaddress]{revtex4}
\usepackage{graphicx}  % needed for figures
\usepackage{subcaption}
\usepackage{mwe}
\usepackage{bm}        % for math
\usepackage{amssymb}   % for math
\usepackage{physics}
\usepackage{float}
\usepackage{verbatim}
\usepackage[dvipsnames]{xcolor}
\usepackage{tikz}

\newcommand{\tikzcircle}[2][red,fill=red]{\tikz[baseline=-0.5ex]\draw[#1,radius=#2] (0,0) circle ;}
\definecolor{green1}{RGB}{51,255,51}

\begin{document}

\title{Graph States as a Resource for Quantum Metrology}
\author{Nathan Shettell}
\affiliation{Laboratoire d’Informatique de Paris 6, CNRS, Sorbonne Université, 4 place Jussieu, 75005 Paris, France}
\author{Damian Markham}
\affiliation{Laboratoire d’Informatique de Paris 6, CNRS, Sorbonne Université, 4 place Jussieu, 75005 Paris, France}
\date{\today}

\begin{abstract}
By using highly entangled states, quantum metrology guarantees precision impossible with classical measurements. 
Unfortunately such states can be very susceptible to noise, and it is a great challenge of the field to maintain quantum advantage in realistic conditions.
In this study we investigate the practicality of graph states for quantum metrology. 
Graph states are a natural resource for much of quantum information, and here we characterize their quantum Fisher information (QFI) for an arbitrary graph state. We then construct families of graph states which approximately achieves the Heisenberg limit, we call these states bundled graph states. We demonstrate that bundled graph states maintain a quantum advantage after being subjected to iid dephasing or finite erasures. This shows that these graph states are good resources for robust quantum metrology. We also quantify the number of $n$ qubit stabilizer states that are useful as a resource for quantum metrology.
\end{abstract}

%see here that they offer some 

%\pacs{}
\maketitle

Quantum metrology describes the framework for estimation strategies which surpass the precision limit of classical strategies \cite{toth2014,giovannetti2006,giovannetti2004,demkowicz2012}. A classical estimation strategy can be emulated by using non-entangled single qubit quantum states to estimate an unknown parameter $\theta$. 
The highest precision we can achieve in this way is such that the mean squared error, $\Delta \theta^2$, scales inversely with the number of quantum states ($n$); $\Delta \theta^2 \geq 1/n$. Using an $n$ qubit entangled state, it is possible to gain a quadratic advantage in precision; $\Delta \theta^2 \geq 1/n^2$, otherwise known as the Heisenberg limit (HL).

Given a quantum resource, $\rho$, the highest achievable precision attainable is bounded by $\mathcal{Q}(\rho)$, the quantum Fisher information (QFI): $\Delta \theta ^2 \geq 1/\mathcal{Q}(\rho)$. There are many different proposals for which quantum states make ideal candidates for quantum metrology, including, but not limited to, spin squeezed states \cite{gross2012}, 2D cluster states \cite{friis2017}, and symmetric states  \cite{oszmaniec2016,Yingkai2019}. A fundamental issue in this area is how to tolerate noise \cite{nichols2016,albarelli2019}. For example, the GHZ state is the canonical, optimal, resource for quantum metrology \cite{toth2014,giovannetti2004,hyllus2012}, but it is very fragile to noise, losing all quantum advantage if only one system is lost \cite{ma2011,zhang2013}.
In this study we explore the practicality of graph states as a resource for quantum metrology.

Graph states are incredibly useful resources with applications spanning many quantum information processes including cryptography \cite{Markham2008,qian2012}, quantum networks \cite{pirker2018,meignant2018}, computation \cite{Raussendorf2001,nielsen2006}, and quantum error correction \cite{Schlingemann2001,hein2006}.
Furthermore, they can be implemented using different techniques, such as, ion traps \cite{barreiro2011}, super conducting qubits \cite{Song2017},  NV centers \cite{cramer2016} and discrete \cite{clark2005,adcock2018,gu2019} and continuous variable \cite{van2007,walschaers2018} optics.
Here we show that they can also be effective resources for practical quantum metrology in the presence of noise.

In our study we consider the canonical case of phase estimation, where an unknown phase $\theta$ is encoded using non-interacting Hamiltonians
\begin{equation}
 U_\theta = \exp(-i \frac{\theta}{2} \sum_{j=1}^n X_j ),
\end{equation}
where $X_j$ is the Pauli $X$ operator acting on the $j$th qubit. In this scenario the QFI for an arbitrary quantum state $\rho=\sum_j \lambda_j \dyad{j}$ is  \cite{petz2011,hyllus2012}
\begin{equation}
\label{eqn:QFI1}
\mathcal{Q}(\rho)=\frac{1}{2} \sum_{\substack{j,k \\ \lambda_j+\lambda_k \neq 0}} \frac{(\lambda_j-\lambda_k)^2}{\lambda_j+\lambda_k} \big| \langle j |\sum_{i=1}^n X_i | k \rangle \big|^2.
\end{equation}
Which can be simplified for pure states $\psi$
\begin{equation}
\label{eqn:QFI2}
\mathcal{Q}(\psi) = \sum_{i,j=1}^n \Tr (X_iX_j \psi) - \Tr(X_i \psi)\Tr(X_j \psi).
\end{equation}

An $n$ qubit stabilizer state, $\psi$, is defined to be the unique $+1$-eigenvalue of $n$ independent Pauli operators $g_1,\ldots,g_n$
\begin{equation}
\label{eqn:Stab}
\psi = 2^{-n} \prod_{i=1}^n (g_i + \mathbb{I} ) = 2^{-n} \sum_{S \in \mathcal{S}} S,
\end{equation}
where $\mathcal{S}$ is the stabilizer group generated by $g_1,\ldots,g_n$,  containing all Pauli operators $S$ which stabilize $\psi$ ($S\psi=\psi$). Aaronson and Gottesman \cite{aaronson2004} compute the number of $n$ qubit stabilizer states, $N_n$, by counting the number of ways $n$ independent generators can be chosen from the Pauli group. It is clear from Eq.~(\ref{eqn:QFI2}) that if the generators are chosen from the Pauli group such that there are no stabilizers of the forms $\pm X_i$ or $-X_iX_j$, then the QFI of the stabilizer state is equal to the number of stabilizers of the form $X_iX_j$. Defining $k=n^{1-\epsilon/2}$, we show in Appendix B that there are at least
\begin{equation}
\label{eqn:qty}
\sum_{j=k}^{n} \binom{n-1}{j-1} 2^{j} N_{n-j} \geq \binom{n-1}{k-1} 2^{k} N_{n-k}
\end{equation}
stabilizer states with a QFI of at least $n^{2-\epsilon}$.

One class of stabilizer states which have no stabilizers of the form $\pm X_i$ or $-X_iX_j$ are graph states with no isolated vertices. Formally, an $n$ qubit graph state $G=(V,E)$ can be defined in correspondence to a simple graph with $n$ vertices $V$ and edges $E$. The corresponding generators are
\begin{equation}
\label{eqn:graph}
g_i= X_i \bigotimes_{j \in N(i)} Z_j,
\end{equation}
where $N(i)$ is the neighbourhood of the $i$th vertex. A graph has no isolated vertices if $|N(i)| \geq 1 \; \forall i$. A graph state is stabilized by $X_iX_j$ if $N(i)=N(j)$, i.e. if the $i$th and $j$th vertices have the same neighbourhood. Hence, the QFI of a graph state $G$ with no isolated vertices is equal to the number of pairs of vertices $(i,j)$ such that $N(i)=N(j)$. By dividing the set of vertices into disjoint sets $\mathcal{V}_1,\mathcal{V}_2,\ldots,\mathcal{V}_l\ldots \subseteq V$, where $|\mathcal{V}_l|=v_l$ we can write
\begin{equation}
\label{eqn:qfiG}
\mathcal{Q} (G) = \sum_l v_l^2.
\end{equation}

Eq.~(\ref{eqn:qfiG}) makes it immediately evident whether a specific graph state is a good resource for quantum metrology. Consider a cluster state, where the associated graph is in the shape of a lattice. Since no two vertices have identical neighbourhoods we get that $v_l=1 \; \forall l$, thus 
\begin{equation}
\label{eqn:qfiCluster}
\mathcal{Q}(G_\text{cluster})=n.
\end{equation}
This result corresponds to what Friis et al.\@ state in \cite{friis2017}: unmodified cluster states do not provide a scaling advantage for quantum metrology. If we instead consider an $n$ qubit star graph, where one central vertex is connected to the remaining $n-1$ exterior vertices, (which is equivalent to the GHZ state up to local unitaries) there are only two disjoint subsets of vertices with identical neighbourhoods: the central vertex in one and every other vertex in the other, thus
\begin{equation}
\label{eqn:qfiStar}
\mathcal{Q}(G_\text{star})=(n-1)^2+1.
\end{equation}

The QFI of graph states can be further improved by performing local Clifford (LC) operations on the graph state first. In graph theory, two vertices $(i,j)$ of a graph satisfying $N(i)\cup \{ i \} = N(j)\cup \{ j \}$ are said to be true twins \cite{cogis2005}; the associated graph state is stabilized by the Pauli operator $Y_iY_j$. Therefore, performing a LC operation $L$
\begin{equation}
LYL^\dagger=X,
\end{equation}
on vertices which has a true twin will increase the number of stabilizers of the form $X_iX_j$. If this is done, the QFI will be equal to the number of pairs of vertices which have identical neighbourhoods or satisfy the true twin condition.
If we instead divide the set of vertices into disjoint sets $\mathcal{U}_1,\mathcal{U}_2,\ldots,\mathcal{U}_m \ldots \subseteq V$, such that $N(i)\cup{i}$ is constant for any vertex $i \in U_m$, where $|\mathcal{U}_m|=u_m$, we obtain that
\begin{equation}
\label{eqn:qfiGLC}
\mathcal{Q} (G^\text{LC}) = \sum_l v_l^2+\sum_m u_m^2 - n.
\end{equation}
The only scenario for a pair of vertices $(i,j)$ to have identical neighbourhoods and satisfy the true twin condition is if $i=j$; this is the reason for the $-n$ term in Eq.~(\ref{eqn:qfiGLC}). The complete graph, where each vertex is connected to every other vertex, satisfies $N(i)\cup\{ i \} = N(j) \cup \{ j \} \; \forall i,j$, thus
\begin{equation}
\label{eqn:qfiCompleteLC}
\mathcal{Q} (G_\text{complete}^\text{LC}) = n^2.
\end{equation}

Ozmenaic et al.\@ showed that most entangled states are not good for quantum metrology \cite{oszmaniec2016}. However, they also showed that most symmetric states are good quantum metrology. Eq.~(\ref{eqn:qfiG}) and (\ref{eqn:qfiGLC}) show a similar result; the more internal symmetry present within a graph state, the higher the QFI.

\begin{figure}
\includegraphics[width=0.45\textwidth]{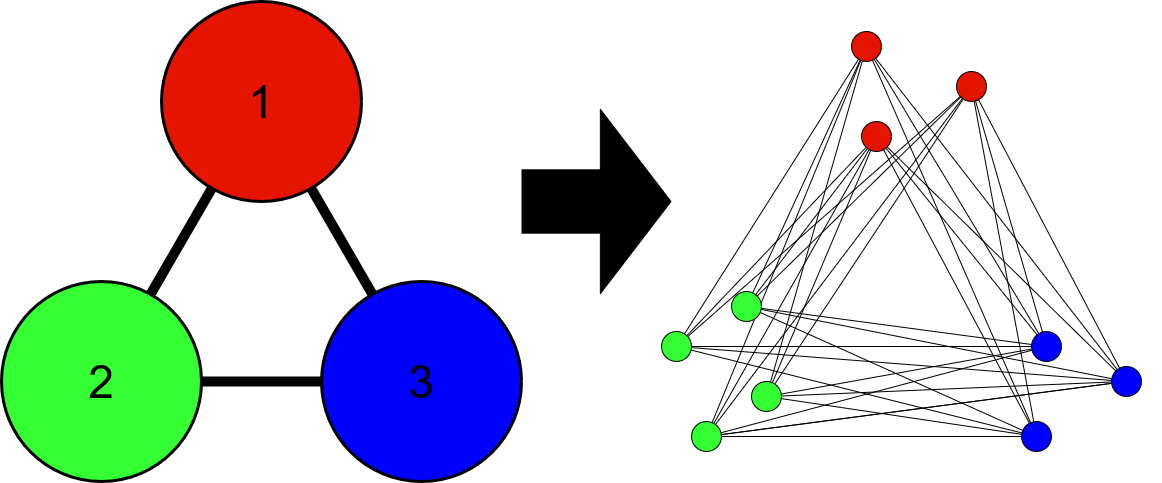}
\begin{center}
\begin{tabular}{  c | c | c | c } 
\textbf{Qubit}& \tikzcircle[black,fill=red]{3.3pt} & \tikzcircle[black, fill=green1]{3.3pt} & \tikzcircle[black,fill=blue]{3.3pt}\\ 
\hline
\textbf{Quantity} & $n_1=3$ & $n_2=4$ & $n_3=3$
\end{tabular}
\end{center}
\caption{Construction of an $n=10$ qubit bundled graph state from a $k=3$ qubit graph state; $\mathcal{Q} = n_1^2+n_2^2+n_3^2 \approx n^{1.5}$.}
\label{fig:BundleExample}
\end{figure}

We provide an easy construction method to transform any $k$ qubit graph state $G=(V,E)$ with no isolated vertices into an $n$ qubit graph state $G_\text{bundle}=(V^\prime,E^\prime)$ $(n > k)$ with many pairs of qubits which have identical neighbourhoods.
\begin{enumerate}
\item Begin with any $k$ qubit graph state $G=(V,E)$ with no isolated vertices.

\item Vertex $i$ is replaced with $n_i$ qubits, labelled $i^{(1)},\ldots,i^{(n_i)}$, such that $\sum_{i=1}^k n_i =n$.

\item If $(i,j) \in E$ then $(i^{(a)},j^{(b)}) \in E^\prime \; \forall a,b$.
\end{enumerate}
The resulting $n$ qubit graph state $G_\text{bundle}=(V^\prime,E^\prime)$ has vertices
\begin{equation}
V^\prime = \{ 1^{(1)}, \ldots, 1^{(n_1)}, \ldots, k^{(1)} , \ldots, k^{(n_k)} \},
\end{equation}
and edges
\begin{equation}
E' = \{ (i^{(a)},j^{(b)}) \; \forall a,b \; | \; (i,j) \in E \}.
\end{equation}
We illustrate the construction of a bundled graph state in Fig.~\ref{fig:BundleExample} by transforming a 3 qubit graph state into an $n=10$ qubit bundled graph state.

\begin{figure*}[ht!]
	\centering
	\begin{subfigure}[b]{0.44\textwidth}
		\centering
		\raisebox{-0.5\height}{\includegraphics[width=\textwidth]{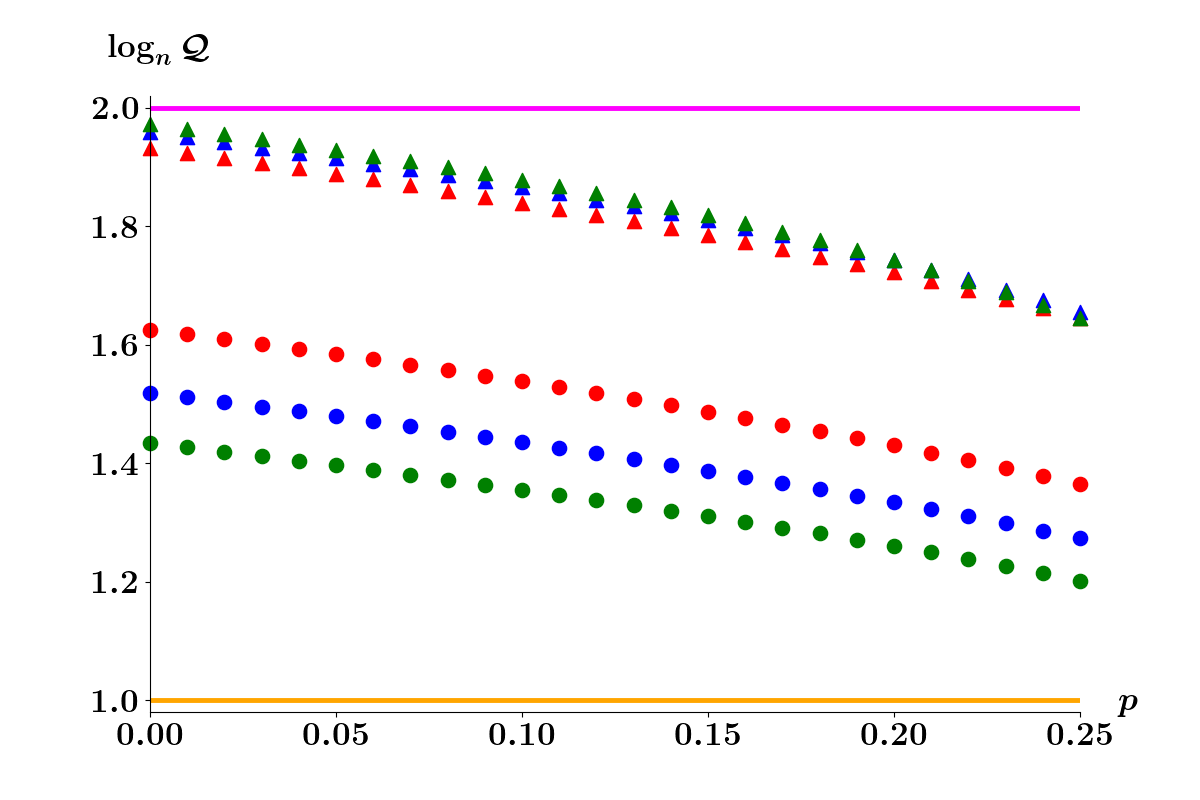}}
		\caption{Bundled Graphs subjected to iid Dephasing}
	\end{subfigure}
	\hfill
	\begin{subfigure}[b]{0.44\textwidth}
		\centering
		\raisebox{-0.5\height}{\includegraphics[width=\textwidth]{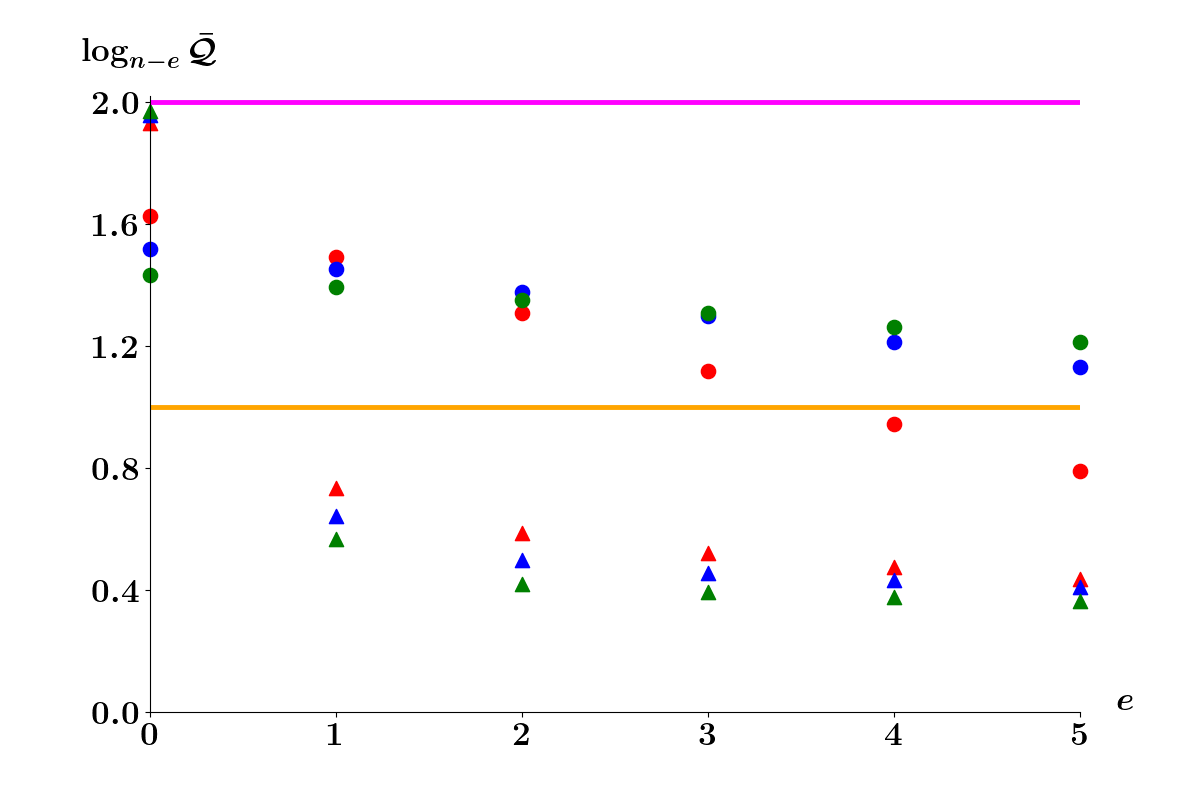}}
		\caption{Bundled Graphs subjected to $e$ Erasures}
	\end{subfigure}
	\hfill
	\begin{subfigure}[b]{0.1\textwidth}
		\centering
		\raisebox{0.4\height}{\includegraphics[width=\textwidth]{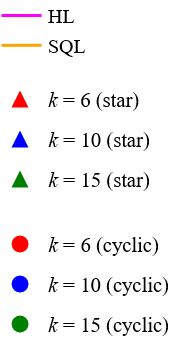}}
	\end{subfigure}
	\caption{Robustness of $n=120$ qubit bundled graphs subjected to iid dephasing (a) and a small number of erasures (b). In each scenario, the graphs are divided into $k$ bundles of $n/k$ qubits. For small dephasing probability $p$, we observe that $\log_n \mathcal{Q}$ decreases linearly, which is expected from Eq.~(\ref{eqn:bundledephasing}). Regarding erasures, we see that bundled cyclic graph states retain a quantum advantage after a small number ($\leq3$) of erasures. In contrast, the QFI of bundled star graphs fall below the SQL after a single erasure.}
	\label{fig:RobustnessPlots}
\end{figure*}

We chose the term bundled graph states because the qubits are divided into bundles of vertices $i^{(1)}, \ldots, i^{(n_i)}$ which all share a common neighbourhood. Using Eq.~(\ref{eqn:qfiG}) to compute that
\begin{equation}
\label{eqn:qfibundle}
\mathcal{Q} (G_\text{bundle}) \geq \sum_{i=1}^{k} n_i^2 \geq \frac{n^2}{k} = n^{2-\log_n k}.
\end{equation}
Note that the shape of graph effects the usefulness of the bundled graph state counterpart. An $n$ qubit bundled cyclic graph divided into $k$ bundles of $n/k$ qubits has a QFI of
\begin{equation}
\mathcal{Q}(G_{\text{bundled cyclic}})=\frac{n^2}{k}.
\end{equation}
Hence, the QFI decreases with the number of bundles when $n$ is kept constant. However, an analogous bundled star graph has a QFI of
\begin{equation}
\mathcal{Q}(G_{\text{bundled star}})=\frac{n^2}{k^2}+\big( n-\frac{n}{k} \big)^2.
\end{equation}
Here the QFI increases as the number of bundles increases. This is because a star graph has a naturally high QFI, and bundling qubits together has a negative impact. Both of these effects are illustrated in Fig.~\ref{fig:RobustnessPlots}.

Next we explore the robustness of graph states subjected to two different noise models. First we investigate the robustness of graph states subjected to independent and identically distributed (iid) dephasing. This transforms a graph state $G$ into a mixed state
\begin{equation}
G \rightarrow G^\text{dephasing}=\sum_{\vec{k}} p^k (1-p)^{n-k} Z_{\vec{k}} G Z_{\vec{k}},
\label{eqn:GraphDephasing}
\end{equation}
where $p$ is the dephasing probability. A closed form expression for $\mathcal{Q}(G^\text{dephasing})$ is derived in Appendix C. We denote $\mathcal{N}_l$ to be the shared neighbourhood of vertices in the set $\mathcal{V}_l$. If the size of the shared neighbourhoods, $N_l=|\mathcal{N}_l|$, are large enough and the dephasing probability is small enough such that
\begin{equation}
(2p(1-p)+1/2)^{N_l} \approx 0
\label{eqn:DephasingApprox}
\end{equation}
then we can approximate that
\begin{equation}
\mathcal{Q}(G^\text{dephasing}) \approx (1-2p)^2 \mathcal{Q}(G) + 4np(1-p).
\end{equation}
Substituting the above in Eq.~(\ref{eqn:qfibundle}) 
\begin{equation}
\label{eqn:bundledephasing}
\begin{split}
\mathcal{Q}(G^\text{dephasing}_\text{bundle}) & \geq (1-2p)^2 \frac{n^2}{k} \\
& = n^{2-\log_n k - \frac{4}{\ln n} p + \mathcal{O}(p^2) }.
\end{split}
\end{equation}
In Fig.~\ref{fig:RobustnessPlots} we see that bundled star graph states and bundled cyclic graph states surpass the SQL for $p \leq 0.2$. Additionally, the subplot of the bundled star graph subjected to iid dephasing demonstrates the viability of the approximation in Eq.~(\ref{eqn:DephasingApprox}). The set of qubits all connected to the central node has a neighbourhood size of $n/k$ and for intermediate values of $p \in [0.1,0.2]$ the term
\begin{equation}
(2p(1-p)+1/2)^{\frac{n}{k}}
\end{equation}
becomes non-negligible when $k$ is large; hence a difference in slope among the family of curves.

The second noisy system we explore is the robustness of graph states subjected to a small number of erasures. In Appendix D we derive a closed form expression for the QFI of a graph state with no isolated vertices subjected to erasures occurring at vertices $y_1,\ldots,y_e$. The general form is extremely dependent on the shape of the graph as well as the choice of $y_1,\ldots,y_e$. To obtain any sort of meaningful value to quantify the robustness we compute $\bar{\mathcal{Q}}$; the average QFI of the system over all $\binom{n}{e}$ permutations of losing $e$ qubits. 

For a general graph $G$, one can show that for a single erasure
\begin{equation}
\label{eqn:1loss}
\bar{\mathcal{Q}}(G^\text{single erasure}) = \sum_l v_l^2 \frac{n-v_l-N_l}{n}+\sum_l v_l \frac{N_l}{n}.
\end{equation}
Notice that the quadratic term is affected by both the size of $\mathcal{V}_l$ as well as the size $\mathcal{N}_l$, indicating that a single erasure propagates to all other qubits with the identical neighbourhood as well as its neighbourhood. In Fig.~\ref{fig:RobustnessPlots}, the QFI of a bundled star graph is less than the standard quantum limit after a single erasure. This is because $n-N_l-v_l=0$ for both the central bundle as well as the set of extremal qubits. However, the QFI of bundled cyclic graphs of different proportions retain a quantum advantage (on average) after a small number erasures, since the quadratic term is non-vanishing. Explicit formulas for the QFI of the bundled star graph and bundled cyclic graph subjected to $e$ erasures can be found in Appendix D.

We have shown that certain graph states have a QFI which surpasses the classical limit. To realize a precision of $\Delta \theta^2=1/\mathcal{Q}$ a POVM which maximizes the Fisher information must be made (see Appendix A). Generally, this POVM is dependent on the unknown parameter \cite{braunstein1994,chiribella2004} or is highly entangled, making the ideal measurement infeasible. The obvious question is: can a precision of $\Delta \theta^2 = 1/\mathcal{Q}(G)$ be achieved with single qubit measurements when using a graph state $G$ as a resource?

If the following conditions are satisfied then we can achieve the desired precision:
\begin{enumerate}

\item The phase we are trying to estimate is small.

\item There exists a stabilizer, $S_M$, for the graph which consists entirely of $Y$ and $Z$ operators.

\end{enumerate}
The desired precision can be attained by measuring in the $S_M$ basis
\begin{equation}
\label{eqn:expvalS}
\begin{split}
\expval{S_M} &= \Tr (e^{i \frac{\theta}{2} \sum_{i=0}^n X_i} S_M e^{-i \frac{\theta}{2} \sum_{i=0}^n X_i} G) \\
&= \Tr (e^{i \theta \sum_{i=0}^n X_i} G) \\
&= \sum_{m=0}^\infty \frac{(i \theta)^m}{m!} \Tr ( ( \sum_{i=0}^n X_i )^m G ) \\
&= 1- \frac{\theta^2}{2}\mathcal{Q}(G) + \mathcal{O}(\theta^3). \\
\end{split}
\end{equation}
Using the error propagation formula
\begin{equation}
\label{eqn:precision}
\Delta \theta^2 = \frac{\Delta S_M^2}{|\partial_\theta \expval{S_M}|^2} = \frac{\theta^2 \mathcal{Q}(G)+\mathcal{O}(\theta^3)}{\theta^2 \mathcal{Q}(G)^2+\mathcal{O}(\theta^3)} \approx \frac{1}{\mathcal{Q}(G)}.
\end{equation}
Measuring a small phase is naturally the regime where quantum metrology is most interesting. The second condition above is always satisfied for any bundled star graph, and is satisfied for bundled cyclic graphs when the number of bundles is a multiple of four. We thus see that even with fixed, local measurements, a precision scaling of $\Delta \theta^2 = 1/\mathcal{Q}$ can be achieved for these states.

In the scenario in which a graph state $G$ does not have a stabilizer consisting entirely of $Y$ and $Z$ operators, we can still achieve a precision of $\Delta \theta^2 = 1/\mathcal{Q}(G)$ by using a graph state with one additional qubit, $G^+$. First we find a stabilizer $S$ such that if vertex $i$ and $j$ have identical neighbourhoods then the Pauli operator in the $i$th and $j$th position of $S$ are either both $Z$ or $Y$ or are both $X$ or $\mathbb{I}$. We denote the set of vertices where the Pauli operator of $S$ is $X$ or $\mathbb{I}$ by $C_S$. Next, we append an additional vertex to $G$ and connect it to all of the vertices in the set $C_S$. This is equivalent to adding a new generator
\begin{equation}
g_{n+1}=X_{n+1} \bigotimes_{j \in  C_S} Z_j.
\end{equation}

Repeating the same computation in  Eq.~(\ref{eqn:expvalS})~and~(\ref{eqn:precision}), we achieve a precision of $\Delta \theta^2 = 1/\mathcal{Q}(G)$ by measuring $G^+$ in the basis of $\tilde{S}_M=g_{n+1}S$. This is a result of the fact that if $X_iX_j$ stabilizes $G$ then it also stabilizes $G^+$ and that $\tilde{S}_M$ contains only Pauli $Y$ and $Z$ operators with the exception of the $n+1$ vertex.

\bigskip

In this study we have presented families of graph states which achieve better than classical scaling in precision, and can be robust against iid dephasing and a small number of erasures. 
These results compare favorably with other resource states for tolerating noise \cite{oszmaniec2016,Yingkai2019}. One may also consider broader noise models, for example collective dephasing \cite{knysh2014}, however, this is beyond our current scope, and for realistic scenarios we do not expect big changes. Another approach is to feed forward quantum error correction strategy, however this can greatly complicate implementation \cite{zhou2018,demkowicz2017,sekatski2017}.
We have also found an expression for the QFI of any graph state.
A main advantage of graph states is that they are already a fundamental resource across quantum information. As such we inherit all the benefits of the work that has gone into their generation \cite{barreiro2011,Song2017,cramer2016,clark2005,
adcock2018,gu2019,van2007,walschaers2018} and distribution over quantum networks \cite{pirker2018,meignant2018}, as well as the flexibility and potential for integration into more elaborate tasks where sensing might be a subroutine.
In this sense graph states make a natural choice for integrating quantum sensing into future quantum networks, and our work demonstrates their capacity to do so.

\textbf{Acknowledgments.} We acknowledge support of the ANR through the ANR-17-CE24-0035 VanQuTe.

\bibliography{GraphStateMetrology_aXv_NS}

\appendix

\onecolumngrid

\setcounter{equation}{0}
\renewcommand\theequation{A.\arabic{equation}}
\section{Appendix A: Fundamentals of Estimation Theory}

Consider an observable random variable $X$ which is dependent on some unknown parameter $\theta$. The goal is to construct an estimate $\hat{\theta}$ based off of the observed outcomes $x_1,\ldots,x_n$. Since $\hat{\theta}$ is dependent on $X$, it is a random variable as well and different moments of $\hat{\theta}$ can be computed. An estimator is unbiased if
\begin{equation}
\langle \hat{\theta} \rangle = \sum_{\vec{x}} p(\vec{x} | \theta) \hat{\theta}(\vec{x}) = \theta.
\end{equation}
Here $p(\vec{x} | \theta)$ is the probability of observing outcome $\vec{x}=x_1,\ldots,x_n$, otherwise known as the likelihood function. For any unbiased estimator, the variance can be bounded below via the Cramer-Rao bound \cite{petz2011,van2004}
\begin{equation}
\Delta \hat{\theta}^2 \geq \frac{1}{\mathcal{I}(X,\theta)}.
\end{equation}
Where $\mathcal{I}(X,\theta)$ is the Fisher information
\begin{equation}
\mathcal{I}(X,\theta) = \langle (\partial_\theta \log p(X|\theta))^2 \rangle = \sum_{\vec{x}} \frac{\big( \partial_\theta p(\vec{x}|\theta) \big)^2}{p(\vec{x}|\theta)}.
\end{equation}
In a scenario where the unknown parameter is encoded into a quantum state $\rho \rightarrow \rho_\theta$, our observable is dependent on the choice of POVM $\{ E_k \}$ and the probability of observing the $k$th outcome is $\Tr (E_k \rho_\theta)$. The Fisher information can be written as
\begin{equation}
\mathcal{I}(\{ E_k \},\theta) = \sum_{k} \frac{\big( \partial_\theta \Tr (E_k \rho_\theta) \big)^2}{\Tr (E_k \rho_\theta)}.
\end{equation}
We now define a new quantity, the quantum Fisher Information (QFI), which is the Fisher information maximized over all POVM's \cite{petz2011}
\begin{equation}
\label{eqnA:QFIdef}
\mathcal{Q} = \max_{\{ E_k \}} \; \mathcal{I}( \{ E_k \}, \theta).
\end{equation}
For a simple encoding $\rho_\theta = e^{-i \theta H} \rho e^{i \theta H}$ with $\rho=\sum_j \lambda_j \dyad{j}{j}$ the QFI has a closed form expression \cite{petz2011,hyllus2012}
\begin{equation}
\label{eqnA:QFI1}
\mathcal{Q}=2 \sum_{\substack{j,k \\ \lambda_j+\lambda_k \neq 0}} \frac{(\lambda_j-\lambda_k)^2}{\lambda_j+\lambda_k} \big| \langle j|H|k \rangle \big|^2.
\end{equation}
Which can be simplified further for pure states $\ket{\psi}$
\begin{equation}
\label{eqnA:QFI2}
\mathcal{Q} = 4\Delta H^2 = 4\expval{H^2}{\psi}-4\expval{H}{\psi}^2.
\end{equation}

\setcounter{equation}{0}
\renewcommand\theequation{B.\arabic{equation}}
\section{Appendix B: Lower bound of $\tilde{N}_{n,\epsilon}$}

We begin by defining two sets of Pauli matrices
\begin{equation}
\begin{split}
A &= \{ X_1X_i \; | \; 1 < i \leq n \}, \\
B_k &= \{ A_1 \ldots A_k \; | \; A_i \in \{ Y,Z \} \; \forall i \}.
\end{split}
\end{equation}
Next we construct the following stabilizer group
\begin{equation}
\mathcal{S}= \langle X_1X_2, \ldots, X_1X_k, P,P g_1, \ldots, P g_{n-k} \rangle.
\end{equation}
Where $P \in B_k$ and $g_1, \ldots , g_{n-k}$ are the generators for any $n-k$ qubit stabilizer state. Notice that the stabilizer group of $\mathcal{S}$ does not contain any stabilizer of the form $\pm X_i$ or $-X_iX_j$, thus the QFI is equal to the number of stabilizers of the form $X_iX_j$, which there are $k^2$ by construction. We define $\tilde{N}_{n,\epsilon}$ to be the number of unique ways in which we can choose a set of generators from $A$, a Pauli operator from $B_k$ and generators $g_1, \ldots , g_{n-k}$ such that the constructed stabilizer state has a QFI of at least $n^{2-\epsilon}$. By choosing $k=n^{1-\epsilon/2}$, we can set the following lower bound
\begin{equation}
\begin{split}
\tilde{N}_{n,\epsilon} &\geq \sum_{j=k}^{n} \binom{n-1}{j-1} |B_{j}| N_{n-j} \\
&= \sum_{j=k}^{n} \binom{n-1}{j-1} 2^{j} N_{n-j} \\
& \geq \binom{n-1}{k-1} 2^{k} N_{n-k}.
\end{split}
\end{equation}

\section{Appendix C: QFI of graph states subjected to iid dephasing}
\setcounter{equation}{0}
\renewcommand\theequation{C.\arabic{equation}}

We model an $n$ qubit graph state $G$ undergoing iid dephasing via
\begin{equation}
G \rightarrow  G^\text{dephasing} = \sum_{\vec{k}} p^k(1-p)^{n-k} Z_{\vec{k}} G Z_{\vec{k}},
\end{equation}
where $p$ is the probability that a qubit undergoes a phase flip. This effectively maps the graph state onto the orthonormal basis $\{ Z_{\vec{k}} \ket{G} \}_{\vec{k}}$. We compute the QFI using Eq.~(\ref{eqnA:QFI2})
\begin{equation}
\mathcal{Q}(G^\text{dephasing}) = \frac{1}{2} \sum_{\vec{j},\vec{k}} \frac{\big(\lambda_{\vec{j}}-\lambda_{\vec{k}}\big)^2}{\lambda_{\vec{j}}+\lambda_{\vec{k}}} \big| \bra{G} Z_{\vec{j}} \sum_i X_i Z_{\vec{k}} \ket{G} \big|^2.
\end{equation}
The only non-zero terms in the sum is if $\vec{j}+\vec{k}=\mathcal{N}_l$ for some $l$. We split $\vec{k}$ into three potions, $a$ qubits with a flipped phase from $\mathcal{V}_l$, $b$ qubits with a flipped phase from $\mathcal{N}_l$ and $c$ qubits with a phase flipped from the remaining qubits,
\begin{equation}
\begin{split}
\mathcal{Q}(G^\text{dephasing}) &= \frac{1}{2} \sum_{\vec{j},\vec{k}} \frac{\big(\lambda_{\vec{j}}-\lambda_{\vec{k}}\big)^2}{\lambda_{\vec{j}}+\lambda_{\vec{k}}} \big| \bra{G} Z_{\vec{j}} \sum_i X_i Z_{\vec{k}} \ket{G} \big|^2 \\
&= \frac{1}{2} \sum_l \sum_{\vec{k}} \frac{\big(\lambda_{\vec{k}+\mathcal{N}_l}-\lambda_{\vec{k}}\big)^2}{\lambda_{\vec{k}+\mathcal{N}_l}+\lambda_{\vec{k}}} \big| \bra{G} Z_{\vec{k}+\mathcal{N}_l} \sum_i X_i Z_{\vec{k}} \ket{G} \big|^2 \\
&= \frac{1}{2} \sum_l \sum_{a,b,c=0} \frac{\big(\lambda_{a-b+c+N_l}-\lambda_{a+b+c}\big)^2}{\lambda_{a-b+c+N_l}+\lambda_{a+b+c}} \big( v_l-2a \big)^2 \\
&= \frac{1}{2} \sum_l f(v_l,p) g(N_l,p).
\end{split}
\end{equation}
Where
\begin{equation}
f(v_l,p)=v_l^2 (1-2p)^2+4v_l p(1-p),
\end{equation}
and
\begin{equation}
\begin{split}
g(N_l,p)&=\sum_{j=0}^{N_l} \binom{N_l}{j} \frac{\big( p^{N_l-j} (1-p)^j - p^j (1-p)^{N_l-j} \big)^2 }{p^{N_l-j} (1-p)^j + p^j (1-p)^{N_l-j}} \\
&=2-4p^{N_l}(1-p)^{N_l} \sum_{j=0}^{N_l} \binom{N_l}{j} \frac{p^j(1-p)^j}{(1-p)^{N_l}p^{2j}+p^{N_l}(1-p)^{2j}} \\
&\geq 2-2 \Big( 2p(1-p)+\frac{1}{2} \Big)^{N_l}.
\end{split}
\label{eqnC:iid}
\end{equation}
From equation Eq.~(\ref{eqnC:iid}) we see that $g\approx 2$ when $(2p(1-p)+1/2)^{N_l} \approx 0$; this is illustrated in Fig.~\ref{fig:ExampleGraph}. Using this approximation, the QFI of $G^\text{dephasing}$ can be written as
\begin{equation}
\label{eqn:qfiDephasing}
\mathcal{Q}(G^\text{dephasing}) \approx \sum_l f(v_l,p) = (1-2p)^2 \mathcal{Q}(G) + 4np(1-p).
\end{equation}

\renewcommand\thefigure{C.\arabic{figure}}

\begin{figure}
\includegraphics[width=0.75\textwidth]{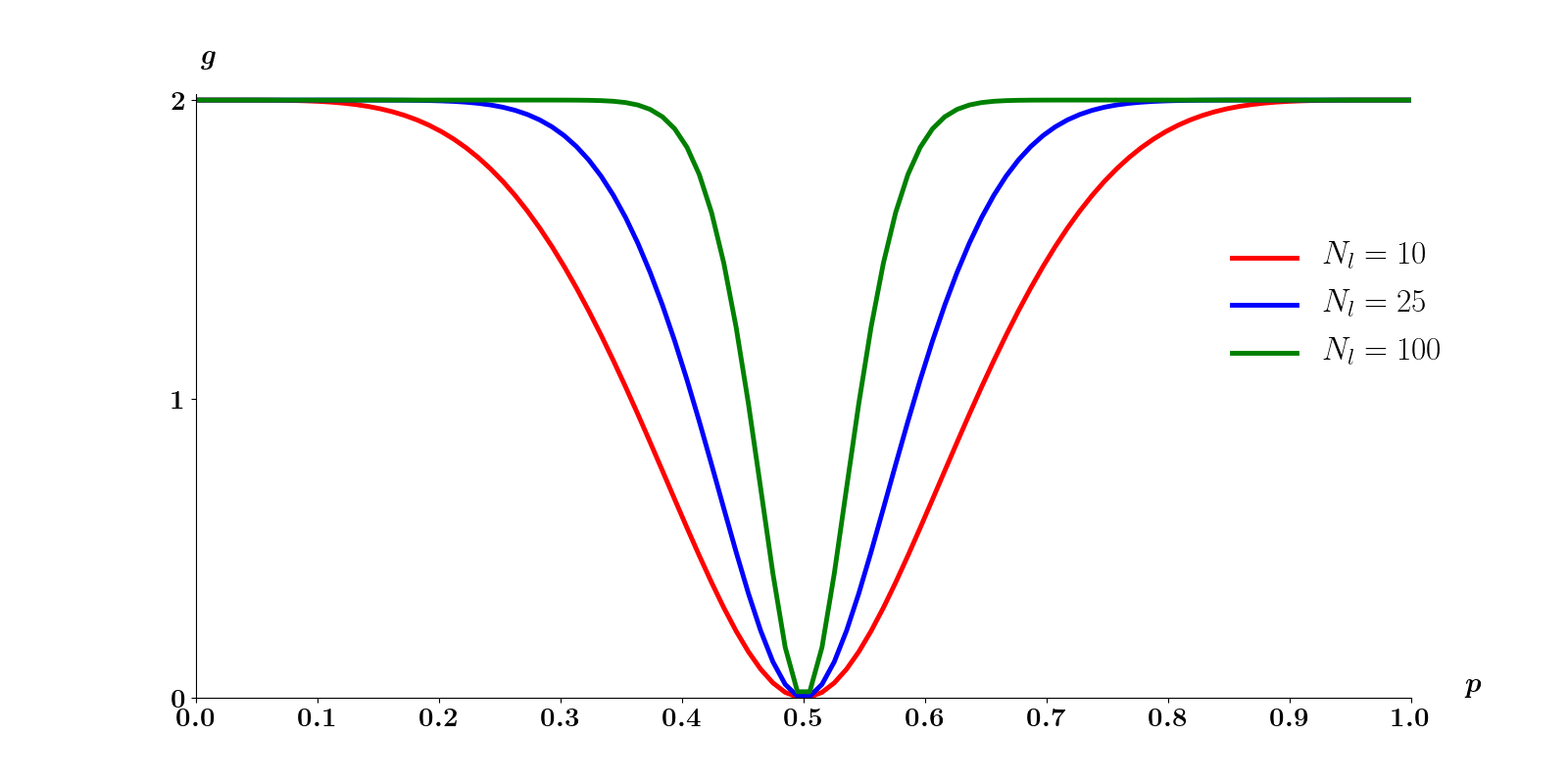}
\caption{Plot of Eq.~(\ref{eqnC:iid}) using varying values of $N_l$. Regardless of the value of $N_l$, $g=0$ when $p=1/2$, this is because $G^\text{dephasing}$ is the maximally mixed state at $p=1/2$, which is useless for quantum metrology.}
\label{fig:ExampleGraph}
\end{figure}

\section{Appendix D: QFI of graph states subjected finite erasures}
\setcounter{equation}{0}
\renewcommand\theequation{D.\arabic{equation}}

We model an $n$ qubit graph state $G$ subjected to finite erasures $\vec{y}=\{ y_1,\ldots,y_e \}$ via
\begin{equation}
G \rightarrow  G_{\vec{y}} = \Tr_{\vec{y}} G.
\end{equation}
This maps $G$ into an equally weighted mixed state
\begin{equation}
G_{\vec{y}} = 2^{-|L_{\vec{y}}|} \sum_{\vec{j} \subseteq L_{\vec{y}} } Z_{\vec{j}} \dyad{G} Z_{\vec{j}}.
\end{equation}
Where the set $L_{\vec{y}}$ is the set which contains all of the lost qubits as well as all of their respective neighbourhoods
\begin{equation}
L_{\vec{y}}=\bigcup_{i=1}^e \{ y_i \} \cup N(y_i).
\end{equation}
Using Eq.~(\ref{eqnA:QFI2}) we compute the QFI of the new state
\begin{equation}
\begin{split}
\mathcal{Q}(G_{\vec{y}}) &= \frac{1}{2} \sum_{\vec{j},\vec{k}} \frac{\big(\lambda_{\vec{j}}-\lambda_{\vec{k}}\big)^2}{\lambda_{\vec{j}}+\lambda_{\vec{k}}} \big| \bra{G} Z_{\vec{j}} \sum_i X_i Z_{\vec{k}} \ket{G} \big|^2 \\
&= \frac{1}{2} \sum_l \sum_{\vec{k}} \frac{\big(\lambda_{\vec{k}+\mathcal{N}_l}-\lambda_{\vec{k}}\big)^2}{\lambda_{\vec{k}+\mathcal{N}_l}+\lambda_{\vec{k}}} \big| \bra{G} Z_{\vec{k}+\mathcal{N}_l} \sum_i X_i Z_{\vec{k}} \ket{G} \big|^2. \\
\end{split}
\end{equation}
Note that $\lambda_{\vec{k}+\mathcal{N}_l}-\lambda_{\vec{k}}=0$ if $\vec{k} \subseteq L_{\vec{y}}$ and $\vec{k}+\mathcal{N}_l \subseteq L_{\vec{y}}$. Regardless of $\vec{k}$, this only occurs  if $\mathcal{N}_l \subseteq L_{\vec{y}}$. In the scenario where $\mathcal{N}_l \nsubseteq L_{\vec{y}}$ the sum over $\vec{k}$ depends on whether $\mathcal{V}_l \subseteq L_{\vec{y}}$ or $\mathcal{V}_l \nsubseteq L_{\vec{y}}$. We can thus simplify the above expression 
\begin{equation}
\label{eqn:erasures1}
\mathcal{Q}(G_{\vec{y}}) = \sum_l h_l(\vec{y}),
\end{equation}
where
\begin{equation}
\label{eqn:erasures2}
h_l(\vec{y}) =
  \begin{cases}
   v_l^2 & \text{if } \mathcal{N}_l \nsubseteq L_{\vec{y}} \text{ and } \mathcal{V}_l \nsubseteq L_{\vec{y}} \\
    v_l & \text{if } \mathcal{N}_l \nsubseteq L_{\vec{y}} \text{ and } \mathcal{V}_l \subseteq L_{\vec{y}} \\
    0 & \text{otherwise } \\
  \end{cases}.
\end{equation}

Eq.~(\ref{eqn:erasures1}) is completely general and thus, it is difficult to understand the repercussions of erasures without specifying the shape of the graph and the erasure sites $\vec{y}$. To obtain To obtain any sort of meaningful value to which quantifies the robustness, we compute $\mathcal{Q}$; the average QFI of the system over all $\binom{n}{e}$ permutations of losing $e$ qubits. For an arbitrary graph $n$ qubit graph $G$, we find that the QFI after a single erasure is
\begin{equation}
\label{eqn:1loss}
\bar{\mathcal{Q}}(G^\text{single erasure}) = \sum_l v_l^2 \frac{n-N_l-v_l}{n}+\sum_l v_l \frac{N_l}{n}.
\end{equation}
Additionally, we can find closed form expressions of $\bar{\mathcal{Q}}$ for multiple erasures for graphs with simple structures. For example, $n$ qubit bundled star graphs with $k$ bundles of $j=n/k$ qubits subjected to $1\leq e \leq n$ erasures
\begin{equation}
\label{eqn:lossStar}
\bar{\mathcal{Q}}(G^{e\text{ erasures}}_\text{bundled star}) = \frac{\binom{n-j}{e}}{\binom{n}{e}} j+\frac{\binom{j}{e}}{\binom{n}{e}} (n-j).
\end{equation}
As well as $n$ qubit bundled cyclic graphs with $k$ bundles of $j=n/k$ qubits subjected to $e<2j$ erasures
\begin{equation}
\label{eqn:lossCyclic}
\bar{\mathcal{Q}}(G^{e\text{ erasures}}_\text{bundled cyclic}) = \frac{2\binom{n-4j}{e}-\binom{n-5j}{e}}{\binom{n}{e}} nj+\frac{2\binom{n-2j}{e}-\binom{n-3j}{e}-2\binom{n-4j}{e}+\binom{n-5j}{e}}{\binom{n}{e}} n.
\end{equation}

\end{document}